\documentclass[journal]{IEEEtran}

\ifCLASSINFOpdf

\else

\fi

\usepackage{amsthm}
\usepackage{epsfig} 
\usepackage{epstopdf}
\usepackage{amsmath}
\usepackage{amsthm}
\usepackage{comment}
\usepackage{url}
\usepackage{algorithm}
\usepackage{algorithmic}

\usepackage{amssymb}
\usepackage{amsmath}
\usepackage{enumerate}
\usepackage{autobreak}  
\usepackage{multirow} 
\usepackage{booktabs} 
\usepackage{makecell}	
\usepackage{threeparttable} 

\usepackage[none]{hyphenat} 

\usepackage{amsfonts}
\usepackage{bm}
\usepackage{color}
\usepackage{cite}
\usepackage{stfloats}
\usepackage{placeins}
\usepackage{chemarrow}
\usepackage{extarrows}
\allowdisplaybreaks

\theoremstyle{definition}
\newtheorem{theorem}{Theorem}
\newtheorem{lemma}[theorem]{Lemma}
\newtheorem{corollary}[theorem]{Corollary}
\newtheorem{proposition}[theorem]{Proposition}

\theoremstyle{definition}

\newtheorem{remark}{Remark}

\usepackage{comment}
\usepackage{ifthen}
\newboolean{showcomments}
\setboolean{showcomments}{true}
\newcommand{\su}[1]{\ifthenelse{\boolean{showcomments}}
	{ \textcolor{blue}{#1}}{}}

\newcommand{\shen}[1]{\ifthenelse{\boolean{showcomments}}
	{ \textcolor{red}{#1}}{}}

\title{On Over-Conservatism under Low-Probability Extreme Events: A KL-Regularized Wasserstein Distributionally Robust Planning Method}

\makeatletter
\newcommand{\fs@blackruled}{%
  \fs@ruled
  \let\KLRW@ruledpre\@fs@pre
  \let\KLRW@ruledmid\@fs@mid
  \let\KLRW@ruledpost\@fs@post
  \let\KLRW@ruledcapt\@fs@capt
  \def\@fs@pre{{\normalcolor\KLRW@ruledpre}}%
  \def\@fs@mid{{\normalcolor\KLRW@ruledmid}}%
  \def\@fs@post{{\normalcolor\KLRW@ruledpost}}%
  \long\def\@fs@capt##1##2{{\normalcolor\KLRW@ruledcapt{##1}{##2}}}%
}
\floatstyle{blackruled}
\restylefloat{algorithm}
\floatstyle{ruled}
\makeatother

\begin{document}
\graphicspath{{Figures/}}
\newcounter{MYtempeqncnt}

\author{
	Xinda Shen, Yifan Su, \IEEEmembership{Member, IEEE}, Bo Hu, \IEEEmembership{Senior Member, IEEE}, Yanbo Chen, \IEEEmembership{Senior Member, IEEE}, Changzheng Shao, \IEEEmembership{Member, IEEE}, and Tao Wu, \IEEEmembership{Member, IEEE}

    
    \thanks{\textit{(Corresponding author: Yifan Su)}}

	

}

\markboth{Journal of \LaTeX\ Class Files,~Vol.~xx, No.~xx, xx~xxxx}%
{Shell \MakeLowercase{\textit{et al.}}: Bare Demo of I EEEtran.cls for IEEE Journals}

\maketitle

\begin{abstract}
The increasing penetration of renewable energy makes power systems more vulnerable to low-probability extreme events, elevating system planning to a critical role in ensuring reliability. Distributionally robust optimization (DRO) based on the Wasserstein distance is widely used to address this challenge. However, we find that the conventional Wasserstein ambiguity set tends to excessively amplify the probabilities of low-probability extreme events, leading to over-conservative investment decisions. To mitigate this over-conservatism, this paper proposes a novel ambiguity set that modifies the Wasserstein distance with a Kullback-Leibler (KL) divergence regularization term. By leveraging the KL divergence’s sensitivity to relative probability changes, the proposed set effectively curbs the unrealistic magnification of low-probability events. A two-stage distributionally robust planning model built upon this new ambiguity set is developed to optimize the installed capacity of photovoltaic, wind, and energy storage. The complex model is then reformulated as an equivalent exponential cone program with strict theoretical proofs, making it computationally tractable. Numerical results on the PJM 5‑bus and IEEE 118‑bus systems reveal the over-conservatism of standard Wasserstein DRO under low-probability extreme events and validate the effectiveness and efficiency of the proposed planning model and solution algorithm.
\end{abstract}


\section*{Nomenclature}

\subsection{Sets and indexes}
{\raggedright
\begin{IEEEdescription}[
\IEEEusemathlabelsep
\IEEEsetlabelwidth{$\mathcal{Z}_{\mathrm N},\mathcal{Z}_{\mathrm A}$cccc}
]
\setlength{\itemsep}{0.16em}
\setlength{\parsep}{0pt}
\setlength{\topsep}{0.12em}

\item[$\mathcal{B},\mathcal{G},\mathcal{L}$]
Sets of buses, thermal generators, and transmission lines.

\item[$\mathcal{T},\mathcal{K}$]
Sets of time periods and normal representative-scenario
indices.

\item[$\mathcal{Z},\mathcal{Z}_{\mathrm N},\mathcal{Z}_{\mathrm A}$]
Historical, normal, and LPEE scenario sets.

\item[$\Xi,\mathcal{X}$]
Uncertainty support and feasible capacity-decision set.

\item[$i,j,g,t,s,n$]
Indices of buses/support points, generators, time periods,
representative scenarios, and historical scenarios;
$n$ also denotes the number of representative scenarios.

\end{IEEEdescription}
\par}

\subsection{Variables and parameters}
{\raggedright
\begin{IEEEdescription}[
\IEEEusemathlabelsep
\IEEEsetlabelwidth{$C^{LS},C^{Cur},C^{Em}$}
]
\setlength{\itemsep}{0.16em}
\setlength{\parsep}{0pt}
\setlength{\topsep}{0.12em}

\item[$x^R_i,x^{Sp}_i,x^{Se}_i$]
Renewable, storage power, and storage energy capacities.

\item[$p^G_{g,t},p^D_{i,t},p^A_{i,t}$]
Thermal generation, load shedding, and renewable curtailment.

\item[$P^{Sc}_{i,t},P^{Sd}_{i,t},e^S_{i,t}$]
Storage charging power, discharging power, and energy state.

\item[$p^L_{ij,t},\theta_{i,t}$]
Line power flow and voltage angle.

\item[$\xi,\widehat{\zeta}_s$]
Uncertainty realization and representative scenario.

\item[$q_n^W,q_n^S,q_n^R$]
Empirical percentile ranks of wind, solar, and total
renewable energy.

\item[$s_n^R,s_n^J,s_n^L$]
Total-energy, joint wind--solar, and persistence scarcity
indicators.

\item[$\ell_n,\chi_n$]
Longest persistent low-output period and scarcity-severity
index.

\item[$p_s,q_s$]
Empirical and candidate scenario probabilities.

\item[$T_{ij},C_{ij}$]
Transported probability mass and ground transportation cost.

\item[$L_s(x),I(x)$]
Scenario-wise operational loss and investment cost.

\item[$W_c,D_{\mathrm{KL}},\mathcal{D}_{\mathrm{KLRW}}$]
Wasserstein distance, KL divergence, and KLRW measure.

\item[$\rho_{\mathrm{KLRW}},\beta$]
KLRW ambiguity radius and Wasserstein--KL trade-off
coefficient.

\item[$C^R,C^{Sp},C^{Se}$]
Renewable, storage power, and storage energy investment
cost coefficients.

\item[$C^{LS},C^{Cur},C^{Em}$]
Load-shedding, curtailment, and emission cost coefficients.

\item[$P^G,R^G,S^L,X^L$]
Parameter families for thermal generation limits, ramping
limits, line capacities, and reactances.

\item[$\eta^S,\omega^S$]
Storage efficiency and state-of-charge parameter families.

\end{IEEEdescription}
\par}

\section{Introduction}
\subsection{Background}
With the accelerating penetration of renewable energy
sources, planning of wind, photovoltaic (PV), and energy
storage has become essential to ensure reliable and
cost-effective integration of power systems. To account for the uncertainties from renewable energy and load, current power system planning approaches are generally categorized into stochastic programming (SP) \cite{baringo2013risk}, robust optimization (RO) \cite{jabr2013robust}, and
distributionally robust optimization (DRO)
\cite{zare2018distributionally,li2023wasserstein}. Among them, DRO has drawn growing attention due to its ability to bridge the gap between data-driven probabilistic information and worst-case robustness.

As global warming intensifies, low-probability extreme events (LPEEs), e.g., extended wind droughts or prolonged heatwaves, pose increasing threats to security of renewable-dominated power systems.
Weather-driven renewable variability and climate change can jointly affect renewable generation and electricity demand, thereby reshaping resource adequacy risks. In particular, renewable energy droughts, characterized by extended periods of low renewable output, may lead to multi-day power shortfalls
\cite{liu2023climate,zheng2024climate,li2024renewable,bracken2024standardized}. For example,
California instituted rotating electricity outages during an extreme heat storm in summer 2020, whereas the 2021 Texas winter storm triggered severe regional generation failures \cite{caiso2021root,ferc2021winter}. Related studies in China further indicate that prolonged high-load conditions, persistent renewable output deficits, and their concurrence can trigger extreme power shortage events and heighten resource adequacy risks \cite{guo2024extreme}.However, the systematic treatments of LPEEs remain underexplored in current planning models, leaving a clear gap in balancing the risks of LPEEs against investment costs.

This paper focuses on the widely adopted Wasserstein-distance-based DRO model for power system planning, aiming to answer the following questions:
\begin{enumerate}
\item Does the Wasserstein-DRO framework exhibit over-conservatism under LPEEs?

\item If so, how should the ambiguity set be modified to mitigate this over-conservatism?

\item Given that classical solution algorithms no longer apply, how can the resulting model be rendered computationally tractable?
\end{enumerate}

\subsection{Related Work}
Power system planning methods can generally be classified according to how uncertainties in renewable generation and demand are modeled. SP assumes that the probability distributions of uncertainties are known and formulates the planning problem by minimizing the expected cost \cite{baringo2013risk}. However, accurately estimating the underlying probability distributions is often difficult in practice. RO assumes that uncertainties vary within a predefined uncertainty set and optimizes the worst-case objective \cite{jabr2013robust}. Nevertheless, its emphasis on worst-case conditions may lead to overly conservative planning decisions. DRO combines the advantages of SP and RO. It assumes that the exact probability distribution is unknown but belongs to a prescribed ambiguity set \cite{zare2018distributionally,xie2022sizing,li2023wasserstein}. It therefore enhances robustness against distributional misspecification while avoiding the excessive conservatism of conventional robust optimization. Therefore, DRO has attracted increasing attention in power system planning.

Existing studies on DRO-based power system planning have focused on constructing more accurate ambiguity sets to characterize uncertainties in renewable generation and demand. Moment-based ambiguity sets characterize candidate distributions using support and low-order statistical information, such as the mean and covariance \cite{delage2010distributionally}, but do not explicitly account for distances between individual uncertainty realizations. Divergence-based ambiguity sets instead define neighborhoods around nominal distributions, e.g, Kullback–Leibler (KL) divergence characterizes distributional deviations through relative entropy \cite{benTal2013robust,chen2018kl}. Wasserstein DRO (WDRO)
constructs an ambiguity set around the empirical distribution using a transportation-based distance between probability distributions \cite{esfahani2018data}. Wasserstein-based DRO can better exploit the underlying physical and geometric information of uncertainties than DRO based on KL divergence, and has therefore been widely adopted in power system planning \cite{duan2018wasserstein,chen2023wasserstein}.

Within the growing of power system planning research, increasing attention has been paid to LPEEs. For instance, Ref. \cite{kittel2026dunkelflaute} shows that extreme renewable droughts can strongly determine long-duration storage requirements. Existing studies mainly address LPEEs through extreme-scenario identification and restriction. Ref. \cite{zhu2024winddrought} generates extreme wind-drought scenarios beyond historical observations for storage expansion planning, while Ref. \cite{bernecker2026adaptive} endogenously incorporate regional Dunkelflaute events into adaptive robust capacity expansion. In Ref. \cite{lin2024extreme}, extreme wind-power scenarios are identified, and a probability uncertainty set is constructed to limits their absolute probability deviations. Ref. \cite{xie2023sizing}, separately models normal and extreme conditions using Wasserstein ambiguity sets and impose a distributionally robust shortfall-risk constraint under extreme conditions. In Ref. \cite{song2026conditional}, deviations of non-typical scenarios are further restricted to narrow the Wasserstein ambiguity set and reduce conservativeness. Overall, existing studies have increasingly incorporated LPEEs into power-system planning, mainly through direct restrictions on extreme-scenario probabilities or deviations. However, the underlying mechanism through how LPEEs affect conservativeness of distributionally robust planning remains insufficiently understood.

\subsection{Contributions}
The main contributions of this paper are threefold:

\begin{enumerate}
\item \textit{\textbf{Source of LPEE-Induced Over-Conservatism:}} This paper reveals an important source of over-conservatism in Wasserstein-based distributionally robust planning. Although LPEEs have small nominal probabilities, their probabilities can increase disproportionately under the worst-case distribution, causing LPEEs to be overemphasized and leading to overly conservative investment decisions.

\item \textit{\textbf{KL-Regularized Wasserstein DRO (KLRW-DRO) Planning Model:}} 
To suppress the over-conservatism issue, a distributionally robust power system planning model is developed with a novel ambiguity set, which is constructed by combining Wasserstein ambiguity with a KL regularization term. Besides, a bootstrap-based calibration method is further developed to determine the radius of ambiguity set by establishing confidence-level equivalence with the classical Wasserstein ambiguity set.

\item \textit{\textbf{Tractable Exponential-Cone Reformulation:}} A tractable reformulation method is developed for the KLRW-DRO planning problem, where affine policies are employed, followed by an exponential-cone reformulation through the strong duality theory. The equivalence of the dual reformulation is rigorously established through theoretical derivation.

\end{enumerate}

\subsection{Organization}
The remainder of this paper is organized as follows. Section II formulates the two-stage distributionally robust planning problem. Section III analyzes the source of LPEE-induced over-conservatism and develops the proposed KLRW ambiguity set, together with the LPEE-aware reference distribution and bootstrap-based radius calibration. Section IV presents the tractable reformulation of the KLRW-DRO model. Section V reports numerical results on the modified PJM 5-bus and IEEE 118-bus systems. Section VI concludes the paper.

\section{Two-Stage Distributionally Robust Planning Problem Formulation}

In this section, we will develop a two-stage DRO model for the coordinated planning of wind power, PV, and energy storage in the power system. In the first stage, the capacities of wind, PV, and energy storage facilities are determined. In the second stage, after the planning decisions are made, the operation of the power system is optimized under uncertain renewable generation and load. The objective is to determine the optimal sizing decisions that minimize the sum of investment costs and the worst-case expected operational cost over all probability distributions within the ambiguity set. The two-stage DRO planning model is formulated in this section, while the construction of the ambiguity set will be presented in the next section.

\subsection{Stage 1: Planning of Wind, PV, and Energy Storage}

In the first stage, the capacities of renewable generation and energy storage are determined before the realization of renewable generation and load uncertainties. The first-stage planning decision is represented by
\begin{equation}
\label{eq:capacity_decision}
x
=
\{x^R_i,x^{Sp}_i,x^{Se}_i\}_{i\in\mathcal{B}}.
\end{equation}
The corresponding investment cost is
\begin{equation}
\label{eq:investment_cost}
I(x)
=
\sum_{i\in\mathcal{B}}
\left(
C^R_i x^R_i
+
C^{Sp}_i x^{Sp}_i
+
C^{Se}_i x^{Se}_i
\right).
\end{equation}
The capacity decisions are subject to the following bounds:
\begin{subequations}
\label{eq:capacity_bounds}
\begin{gather}
0
\leq
x^R_i
\leq
X^{R,u}_i,
\label{eq:renewable_capacity_bound}
\\
0
\leq
x^{Sp}_i
\leq
X^{Sp,u}_i,
\label{eq:storage_power_capacity_bound}
\\
0
\leq
x^{Se}_i
\leq
X^{Se,u}_i.
\label{eq:storage_energy_capacity_bound}
\end{gather}
\end{subequations}
Once the first-stage planning decision is determined, the capacities remain fixed, and the power system is subsequently operated under the realized renewable generation and load uncertainties.

\subsection{Stage 2: Operation of Power System}
In the second stage, given the first-stage capacity
decision \(x\) and uncertainty realization
\(\xi=(\xi^R,\xi^D)\in\Xi\), representing renewable
availability and load demand uncertainties,
respectively, the operational loss is minimized
over the feasible operating decisions
\(z\in\mathcal{F}(x,\xi)\), as defined by the
following constraints:
\begin{equation}
\label{eq:operation_loss}
\begin{aligned}
L(x,\xi)
=
\min_{z\in\mathcal{F}(x,\xi)} \quad
&
\sum_{t\in\mathcal{T}}
\sum_{i\in\mathcal{B}}
\left(
C^{LS}p^D_{i,t}
+
C^{Cur}p^A_{i,t}
\right)
\\
&
+
C^{Em}
\sum_{t\in\mathcal{T}}
\sum_{g\in\mathcal{G}}
F^G_g p^G_{g,t}\Delta t .
\end{aligned}
\end{equation}

The operation of thermal generators is subject to the generation capacity and ramping limits:
\begin{subequations}
\label{eq:thermal_constraints}
\begin{gather}
P^{G,l}_g
\leq
p^G_{g,t}
\leq
P^{G,u}_g ,
\label{eq:thermal_capacity}
\\
R^{G,l}_g\Delta t
\leq
p^G_{g,t}-p^G_{g,t-1}
\leq
R^{G,u}_g\Delta t .
\label{eq:thermal_ramping}
\end{gather}
\end{subequations}

The available renewable generation is determined by the installed renewable capacity and the realized renewable availability:
\begin{equation}
\label{eq:renewable_output}
P^R_{i,t}
=
x^R_i \xi^R_{i,t}.
\end{equation}

The operation of energy storage is constrained by
\begin{subequations}
\label{eq:storage_constraints}
\begin{gather}
0
\leq
P^{Sc}_{i,t}
\leq
x^{Sp}_i,
\qquad
0
\leq
P^{Sd}_{i,t}
\leq
x^{Sp}_i ,
\label{eq:storage_power}
\\
e^S_{i,t}
=
e^S_{i,t-1}
+
\left(
\eta^{Sc}P^{Sc}_{i,t}
-
\frac{P^{Sd}_{i,t}}{\eta^{Sd}}
\right)\Delta t ,
\label{eq:storage_balance}
\\
\omega^{S,l}x^{Se}_i
\leq
e^S_{i,t}
\leq
\omega^{S,u}x^{Se}_i ,
\label{eq:storage_capacity}
\\
e^S_{i,|\mathcal{T}|}
=
e^S_{i,0}.
\label{eq:storage_cycle}
\end{gather}
\end{subequations}

The transmission network is represented by the DC power-flow model:
\begin{equation}
\label{eq:network_constraints}
p^L_{ij,t}
=
\frac{\theta_{i,t}-\theta_{j,t}}{X^L_{ij}},
\quad
-S^L_{ij}
\leq
p^L_{ij,t}
\leq
S^L_{ij}.
\end{equation}

The nodal power balance is given by
\begin{align}
\label{eq:power_balance}
&
\sum_{j:(j,i)\in\mathcal{L}}
p^L_{ji,t}
+
\sum_{g\in\mathcal{G}_i}
p^G_{g,t}
+
P^R_{i,t}
+
P^{Sd}_{i,t}
+
p^D_{i,t}
-
p^A_{i,t}
\nonumber\\
&=
\sum_{j:(i,j)\in\mathcal{L}}
p^L_{ij,t}
+
P^{Sc}_{i,t}
+
\xi^D_{i,t}.
\end{align}

Load shedding and renewable curtailment satisfy
\begin{subequations}
\label{eq:shedding_curtailment_constraints}
\begin{gather}
0
\leq
p^A_{i,t}
\leq
P^R_{i,t},
\label{eq:curtailment}
\\
0
\leq
p^D_{i,t}
\leq
\xi^D_{i,t}.
\label{eq:shedding}
\end{gather}
\end{subequations}

Accordingly, for each representative uncertainty scenario
\(\widehat{\zeta}_s\), the corresponding second-stage operational loss is
\begin{equation}
\label{eq:scenario_loss}
L_s(x)
=
L(x,\widehat{\zeta}_s).
\end{equation}

Let \(\widehat{\mathbb P}=\sum_{s=1}^{n}
p_s\delta_{\widehat{\zeta}_s}\) denote the reference
distribution, where \(p_s>0\) and \(\sum_s p_s=1\).
For the proposed KLRW ambiguity set, any candidate
distribution \(P\) is supported on the representative
scenarios, as established in
Lemma~\ref{lem:support_reduction}. Thus, the expected
operational loss is
\begin{equation}
\label{eq:expected_operation_loss}
\mathbb{E}_{\xi\sim P}[L(x,\xi)]
=
\sum_{s=1}^{n}q_sL_s(x),
\end{equation}
where \(q_s\geq0\) and \(\sum_s q_s=1\) are the
candidate scenario probabilities under \(P\).
The reference distribution and ambiguity set are
constructed in the next section.

\subsection{Overall Problem}
Combining the first-stage capacity planning decisions with the second-stage operational problem, the two-stage DRO planning model is formulated as
\begin{equation}
\label{eq:dro_capacity_planning}
\min_{x\in\mathcal{X}}
\left\{
I(x)
+
\sup_{P\in\mathcal{B}(\widehat{\mathbb P})}
\mathbb{E}_{\xi\sim P}
\left[
L(x,\xi)
\right]
\right\},
\end{equation}
where \(\mathcal{X}\) denotes the feasible set of the first-stage capacity decisions defined by \eqref{eq:capacity_bounds}, and \(L(x,\xi)\) is the second-stage operational loss defined in \eqref{eq:operation_loss}. The ambiguity set \(\mathcal{B}(\widehat{\mathbb P})\) characterizes the candidate probability distributions of renewable generation and load uncertainties around the reference distribution \(\widehat{\mathbb P}\).

It should be noted that the planning results of a DRO model are inherently dependent on the specification of the ambiguity set. In particular, when the LPEEs are incorporated into the planning model, conventional ambiguity-set formulations may lead to overly conservative planning decisions, potentially resulting in excessive investment. This key issue is how distributional uncertainty is characterized by the ambiguity set, which will be addressed in the next section.

\section{KL-Regularized Wasserstein Ambiguity Set}
In this section, we first investigate the source of over-conservatism induced by the classical Wasserstein ambiguity set when the LPEEs are incorporated into the planning problem. Accordingly, a novel KL-regularized Wasserstein ambiguity set is proposed to better characterize distributional uncertainty while mitigating unnecessary conservatism. In addition, the LPEE-aware reference distribution is constructed, and then a bootstrap-based calibration method is developed to determine the ambiguity-set radius by establishing confidence-level equivalence with the classical Wasserstein ambiguity set.

\subsection{Source of LPEE-Induced Over-Conservatism}

To illustrate the source of LPEE-induced over-conservatism,
we consider a single-bus planning problem, where additional
dispatchable capacity is determined before the uncertain
residual demand is realized. The WDRO model is formulated as
\begin{equation}
\label{eq:motivating_model}
\min_{0\leq x\leq X^{u}}
\left\{
C^{I}x+
\sup_{P\in\mathcal{B}_{\rho}^{W}(\widehat{\mathbb P})}
\mathbb{E}_{\xi\sim P}
\left[L(x,\xi)\right]
\right\}.
\end{equation}

The second-stage operational loss is
\begin{equation}
\label{eq:motivating_operation}
\begin{aligned}
L(x,\xi)
=
\min_{p^{G},p^{D}}\quad
& C^{G}p^{G}+C^{LS}p^{D}
\\
\mathrm{s.t.}\quad
& p^{G}+p^{D}=\xi,
\\
& 0\leq p^{G}\leq K^{0}+x,
\\
& p^{D}\geq0 .
\end{aligned}
\end{equation}
where \(K^{0}\) denotes the existing dispatchable capacity,
and \(\mathcal{B}_{\rho}^{W}(\widehat{\mathbb P})\) is the
Wasserstein ambiguity set with radius \(\rho\).

Set \(K^{0}=1.5\), \(X^{u}=1.5\), \(C^{I}=0.4\),
\(C^{G}=1\), \(C^{LS}=10\), and \(\rho=0.15\).
The two empirical distributions below have the same mean
residual demand of \(1.5\), and the worst-case distributions
are evaluated at the reference capacity \(\bar{x}=0.5\).

\textit{Example 1 (Without LPEE):}
The residual demand takes values
\(\widehat{\zeta}_{1}=1\), \(\widehat{\zeta}_{2}=1.5\),
and \(\widehat{\zeta}_{3}=2\). The empirical distribution
\((0.33,0.34,0.33)\) shifts to the worst-case distribution
\((0.33,0.04,0.63)\).

\textit{Example 2 (With LPEE):}
An LPEE with residual demand \(\widehat{\zeta}_{LPEE}=4\)
is introduced, yielding the empirical distribution
\((0.35,0.34,0.30,0.01)\). The worst-case distribution
becomes \((0.35,0.34,0.225,0.085)\).

\begin{figure}[!t]
\centering
\includegraphics[width=\columnwidth]{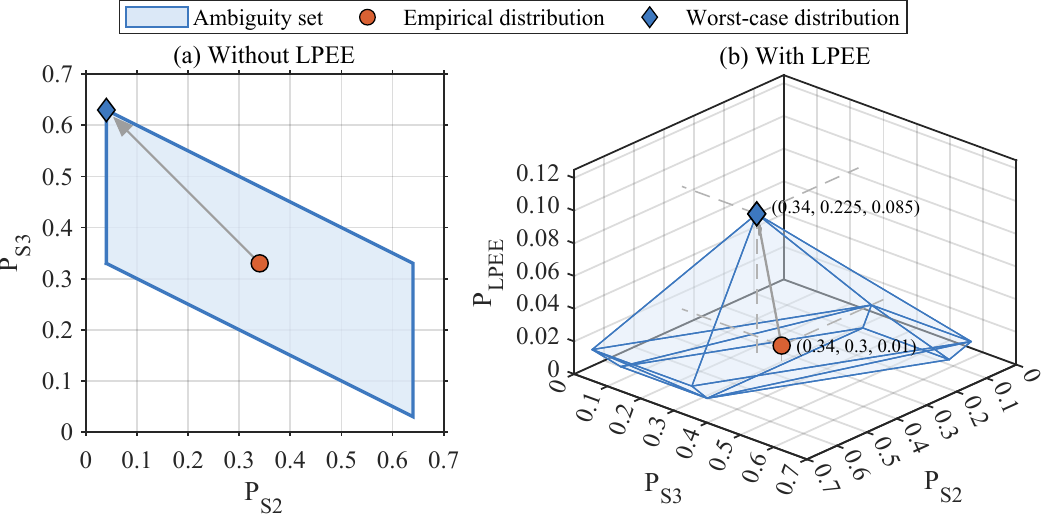}
\caption{Wasserstein ambiguity sets and associated empirical and worst-case distributions: (a) without LPEE and (b) with LPEE.}
\label{fig:motivating_distribution}
\end{figure}

As shown in Fig.~\ref{fig:motivating_distribution}(a)
and (b), the ambiguity sets are represented in the
\((P_{S2},P_{S3})\) and
\((P_{S2},P_{S3},P_{\mathrm{LPEE}})\) probability spaces,
respectively, with \(P_{S1}\) determined by the probability
normalization constraint.

In Example~1, the probability of the most adverse normal
scenario is amplified by a factor of \(1.91\).
In Example~2, the LPEE probability has an \(8.5\)-fold
amplification despite a smaller absolute probability change.
This contrast shows that Wasserstein distance controls
the transportation cost of probability mass but does not
directly restrict relative probability changes, allowing
LPEEs with small empirical probabilities to be
disproportionately amplified in the worst-case distribution.
In Example~2, such probability distortion increases the
optimal additional capacity from \(x^{*}=0.5\) under the
empirical distribution to \(x^{*}=1.5\) under WDRO,
thereby motivating KL regularization to suppress excessive
relative probability deviations toward LPEEs.

\subsection{KL-Regularized Wasserstein Ambiguity Set}

Motivated by the above observation, we introduce a KL divergence regularization term into the Wasserstein distance to restrict excessive relative probability deviations toward LPEEs. The KLRW measure is defined as
\begin{equation}
\label{eq:klrw_measure}
\mathcal{D}_{\mathrm{KLRW}}(P\Vert\widehat{\mathbb P})
:=
(1-\beta)W_c(P,\widehat{\mathbb P})
+
\beta D_{\mathrm{KL}}(P\Vert\widehat{\mathbb P}),
\end{equation}
where \(\beta\in(0,1)\) is the trade-off coefficient between the Wasserstein distance and KL divergence. The Wasserstein component is given by
\begin{equation}
\label{eq:wasserstein_component}
W_c(P,\widehat{\mathbb P})
=
\inf_{\gamma\in\Gamma(P,\widehat{\mathbb P})}
\mathbb{E}_{(\xi,\zeta)\sim\gamma}
\left[
c(\xi,\zeta)
\right],
\end{equation}
where \(\Gamma(P,\widehat{\mathbb P})\) denotes the set of couplings with marginals \(P\) and \(\widehat{\mathbb P}\). Accordingly, the KLRW ambiguity set is defined as
\begin{equation}
\label{eq:klrw_ambiguity_set}
\mathcal{B}_{\rho_{\mathrm{KLRW}}}(\widehat{\mathbb P})
=\bigl\{P\!\in\!\mathcal{M}(\Xi):
\mathcal{D}_{\mathrm{KLRW}}(P\Vert\widehat{\mathbb P})
\!\leq\!\rho_{\mathrm{KLRW}}\bigr\}.
\end{equation}
where \(\rho_{\mathrm{KLRW}}\) is the ambiguity radius. The Wasserstein component characterizes distributional deviations through probability-mass transportation, while the KL component restricts excessive relative probability deviations from the reference distribution.

For a finite-support reference distribution
\(\widehat{\mathbb P}
=\sum_{i=1}^{n}p_i\delta_{\widehat{\zeta}_i}\),
the following result characterizes the support of any distribution in the KLRW ambiguity set.

\begin{lemma}[Support Reduction of the KLRW Ambiguity Set]
\label{lem:support_reduction}
Suppose that \(p_i>0\), \(i=1,\ldots,n\), \(\beta\in(0,1)\), and
\(\rho_{\mathrm{KLRW}}<\infty\). Then every distribution
\(P\in\mathcal{B}_{\rho_{\mathrm{KLRW}}}(\widehat{\mathbb P})\)
satisfies \(P\ll\widehat{\mathbb P}\) and is therefore supported on the empirical atoms.
\end{lemma}

\begin{proof}
See Appendix~\ref{app:support_reduction}.
\end{proof}

By Lemma~\ref{lem:support_reduction}, the distributional problem reduces to probability redistribution over the empirical atoms:
\begin{equation}
\label{eq:worst_case_distribution}
P
=
\sum_{i=1}^{n}
q_i\delta_{\widehat{\zeta}_i},
\qquad
q\in\Delta_n .
\end{equation}
Let
\begin{equation}
\label{eq:ground_cost}
C_{ij}
=
c(\widehat{\zeta}_i,\widehat{\zeta}_j)
=
\left\|
\widehat{\zeta}_i-\widehat{\zeta}_j
\right\|_1 .
\end{equation}
Then the KLRW measure admits the following finite-dimensional form:
\begin{equation}
\label{eq:discrete_klrw}
\begin{alignedat}{2}
\mathcal{D}_{\mathrm{KLRW}}(q\Vert p)
&=
\min_{T\geq0}
\quad
&&
(1-\beta)
\sum_{i=1}^{n}
\sum_{j=1}^{n}
C_{ij}T_{ij}
\\[-0.5ex]
&&&
{}+
\beta
\sum_{i=1}^{n}
q_i\log\frac{q_i}{p_i}
\\[-0.3ex]
&\mathrm{s.t.}
\quad
&&
\sum_{j=1}^{n}T_{ij}=q_i,
\quad
\sum_{i=1}^{n}T_{ij}=p_j .
\end{alignedat}
\end{equation}
Here, \(T_{ij}\) denotes the probability mass transported from empirical scenario \(j\) to worst-case scenario \(i\); its row sums give \(q\), while its column sums equal \(p\).

Revisiting Example~2 with \(\beta=0.97\) and all other
settings unchanged yields the worst-case distribution
\((0.3285,0.3273,0.2963,0.0479)\).
The worst-case LPEE probability increases from its
empirical value of \(1\%\) to \(4.79\%\), compared with
\(8.5\%\) under WDRO, illustrating the suppression of
excessive LPEE probability amplification by KL
regularization.

\begin{remark}[Mechanism for Mitigating LPEE-Induced Over-Conservatism]
The effectiveness of the proposed KLRW ambiguity set stems from the complementary roles of its Wasserstein and KL components. The Wasserstein criterion restricts distributional deviations through probability-mass transportation in the uncertainty space, but it does not explicitly control the relative change in probability assigned to individual scenarios. Consequently, LPEEs may still be disproportionately amplified despite involving only a small amount of probability mass. The KL regularization complements this limitation by penalizing excessive probability deviations relative to the reference distribution, particularly large relative probability changes associated with low-probability events. Therefore, the Wasserstein component preserves robustness against plausible uncertainty deviations, while the KL component suppresses excessive probability concentration toward LPEEs, thereby alleviating LPEE-induced over-conservatism.
\end{remark}

\begin{remark}[Difference from Sinkhorn-Distance-Based DRO]
The proposed KLRW ambiguity set should be distinguished from the Sinkhorn ambiguity set adopted in \cite{zhou2026sinkhorn}, although both involve Wasserstein and KL-related information. Specifically, Sinkhorn distance introduces a KL regularization term on the transport plan, whereas the KLRW measure directly incorporates the KL divergence between the worst-case distribution and the reference distribution. Moreover, the two ambiguity sets are designed to address different limitations of Wasserstein-based DRO. The Sinkhorn ambiguity set mainly aims to alleviate the discreteness of the worst-case distribution, while the KLRW ambiguity set is developed to mitigate LPEE-induced over-conservatism.
\end{remark}

With the KLRW ambiguity set established, the reference
distribution and ambiguity radius remain to be specified.
The following two subsections present their construction
and calibration, respectively.

\subsection{LPEE-Aware Empirical Distribution Construction}

Following the definition of the KLRW ambiguity set, the
reference distribution is constructed while retaining
low-probability extreme events (LPEEs).
Let \(\mathcal{Z}=\{\zeta_n\}_{n=1}^{N_h}\) denote the
historical scenario set obtained by segmenting wind and
solar trajectories into operating scenarios over the
planning horizon. For each scenario \(\zeta_n\), let
\(q_n^W\), \(q_n^S\), and \(q_n^R\in[0,1]\) denote the
empirical percentile ranks of wind, solar, and total
renewable energy, respectively, and let \(\ell_n\) denote
the longest persistent low-output period.
The scarcity indicators are defined as
\begin{equation}
\label{eq:scarcity_indicators}
\left\{
\begin{aligned}
s_n^R &= 1-q_n^R,\\
s_n^J &= \min\{1-q_n^W,\,1-q_n^S\},\\
s_n^L &= \frac{\ell_n}{|\mathcal{T}|}.
\end{aligned}
\right.
\end{equation}
The scarcity-severity index is defined as
\begin{equation}
\label{eq:abnormal_index}
\chi_n
=
\omega_R s_n^R
+
\omega_J s_n^J
+
\omega_L s_n^L,
\end{equation}
where \(\omega_R,\omega_J,\omega_L\geq0\) and
\(\omega_R+\omega_J+\omega_L=1\).
A larger \(\chi_n\) indicates more severe renewable scarcity.

Scenarios in the upper tail of the scarcity-severity index
are classified as LPEEs:
\begin{equation}
\label{eq:abnormal_set}
\mathcal{Z}_{\mathrm A}
=
\left\{
\zeta_n\in\mathcal{Z}:
\chi_n
\geq
\widehat Q_{\alpha}(\chi)
\right\}.
\end{equation}
Here, \(\widehat Q_{\alpha}(\chi)\) denotes the empirical
\(\alpha\)-quantile of the scarcity-severity index.
The remaining scenarios constitute the normal scenario set:
\begin{equation}
\label{eq:normal_set}
\mathcal{Z}_{\mathrm N}
=
\mathcal{Z}\setminus\mathcal{Z}_{\mathrm A}.
\end{equation}

The most severe LPEE scenario is selected as the representative
extreme scenario \(\widehat{\zeta}_{A}\).
The normal scenarios are reduced using hard-balanced
clustering \cite{lin2019balanced}, yielding balanced
normal clusters
\(\{\widetilde{\mathcal{C}}_k\}_{k\in\mathcal K}\).
The historical scenario closest to each cluster center
is selected as the representative normal scenario
\(\widehat{\zeta}_{N,k}\).
The empirical probabilities are
\begin{equation}
\label{eq:empirical_probabilities}
p_{N,k}
=
\frac{|\widetilde{\mathcal{C}}_k|}
{|\mathcal{Z}_{\mathrm N}|+|\mathcal{Z}_{\mathrm A}|},
\qquad
p_A
=
\frac{|\mathcal{Z}_{\mathrm A}|}
{|\mathcal{Z}_{\mathrm N}|+|\mathcal{Z}_{\mathrm A}|}.
\end{equation}
The resulting LPEE-aware empirical distribution is
\begin{equation}
\label{eq:empirical_distribution}
\widehat{\mathbb P}
=
\sum_{k\in\mathcal K}
p_{N,k}\delta_{\widehat{\zeta}_{N,k}}
+
p_A\delta_{\widehat{\zeta}_{A}}.
\end{equation}
Hereafter, its support and probability vector are denoted by
\(\{\widehat{\zeta}_s\}_{s=1}^{n}\) and
\(p=(p_1,\ldots,p_n)\), respectively.

\subsection{Bootstrap-based Radius Calibration}
With the reference distribution constructed, the ambiguity radius remains to be determined. To ensure a consistent level of distributional robustness, we calibrate the KLRW ambiguity radius through bootstrap at the same prescribed confidence level as the classical Wasserstein ambiguity set.

For each bootstrap replication \(b=1,\ldots,B\), a weighted sample is generated according to the empirical probability vector \(p=(p_1,\ldots,p_n)\), yielding a bootstrap empirical distribution \(q^{(b)}\). The corresponding Wasserstein distance and KL divergence from the reference distribution are calculated as
\begin{subequations}
\label{eq:bootstrap_component_calibration}
\begin{align}
d_W^{(b)}
&=
W_c\left(q^{(b)},p\right),
\\
d_{\mathrm{KL}}^{(b)}
&=
D_{\mathrm{KL}}\left(q^{(b)}\Vert p\right).
\end{align}
\end{subequations}

For the classical Wasserstein ambiguity set, the radius at confidence level \(1-\eta\) is given by
\begin{equation}
\label{eq:wasserstein_radius_calibration}
\rho_W
=
Q_{1-\eta}
\left(
\{d_W^{(b)}\}_{b=1}^{B}
\right),
\end{equation}
where \(Q_{1-\eta}\) denotes the empirical \((1-\eta)\)-quantile.

For a prescribed \(\beta\), the KLRW distance associated with the same bootstrap replication is
\begin{equation}
\label{eq:bootstrap_klrw_statistic}
d_{\mathrm{KLRW}}^{(b)}(\beta)
=
(1-\beta)d_W^{(b)}
+
\beta d_{\mathrm{KL}}^{(b)}.
\end{equation}
Accordingly, the KLRW ambiguity radius is calibrated as
\begin{equation}
\label{eq:klrw_radius_calibration}
\rho_{\mathrm{KLRW}}(\beta)
=
Q_{1-\eta}
\left(
\{d_{\mathrm{KLRW}}^{(b)}(\beta)\}_{b=1}^{B}
\right).
\end{equation}
Thus, the Wasserstein and KLRW ambiguity sets are calibrated using the same bootstrap samples and confidence level, providing a consistent basis for comparison. In this study, \(1-\eta=0.95\), and \(\rho_{\mathrm{KLRW}}(\beta)\) is recalibrated whenever \(\beta\) changes.

\begin{algorithm}[!t]
\caption{KLRW Ambiguity-Radius Calibration}
\label{alg:klrw_calib}
\begin{algorithmic}[1]
\REQUIRE Empirical support
\(\{(\widehat{\zeta}_s,p_s)\}_{s=1}^{n}\);
ground-cost matrix \(C\); bootstrap replications \(B\);
confidence level \(1-\eta\); coefficient \(\beta\in(0,1)\).
\ENSURE \(\rho_{\mathrm{KLRW}}(\beta)\).

\FOR{\(b=1,\ldots,B\)}
    \STATE Generate a weighted bootstrap sample according to \(p\).
    \STATE Construct \(q^{(b)}\) and compute
    \(d_W^{(b)}\), \(d_{\mathrm{KL}}^{(b)}\), and
    \(d_{\mathrm{KLRW}}^{(b)}(\beta)\).
\ENDFOR
\STATE Set \(\rho_{\mathrm{KLRW}}(\beta)\) to the empirical
\((1-\eta)\)-quantile of
\(\{d_{\mathrm{KLRW}}^{(b)}(\beta)\}_{b=1}^{B}\).
\RETURN \(\rho_{\mathrm{KLRW}}(\beta)\).
\end{algorithmic}
\end{algorithm}

While the proposed KL-regularized Wasserstein ambiguity set could alleviate the LPEE-induced over-conservatism, the combination of Wasserstein distance and KL divergence significantly complicates the two-stage DRO planning problem. Conventional solution approaches for Wasserstein-and KL-based DRO are derived under their respective ambiguity-set structures and are not directly transferable to the proposed hybrid ambiguity set. This issue motivates the exponential-cone reformulation developed in the next section.

\section{Tractable Reformulation of KLRW-DRO}
\label{sec:reformulation}

With the KLRW ambiguity set and its key parameters specified, we next address the computational tractability of the resulting two-stage planning problem.
This section develops a tractable reformulation of the proposed KLRW-DRO planning model. The affine policy is introduced and the resulting robust operational constraints are reformulated into a finite-dimensional form. Then, the KLRW-DRO operational risk is reformulated through strong duality.

\subsection{Affine Policy and Robust Operational Reformulation}

An affine recourse policy is imposed on the second-stage
operational decisions. Let
\(\xi=(\xi_1,\ldots,\xi_K)^\top\) collect renewable
availability and load demand over the planning horizon.
The policy is
\begin{equation}
\label{eq:affine_policy_compact}
z(\xi)=y+\mathbf{Y}\xi,
\end{equation}
where \(z(\xi)=[p^D_{i,t},\,p^A_{i,t},\,P^{Sc}_{i,t},\,
P^{Sd}_{i,t},\,e^S_{i,t},\,p^L_{ij,t},\,\theta_{i,t},\,
p^G_{g,t}]^\top\) stacks all indexed operational variables,
with dependence on \(\xi\) implicit; \(y\) and
\(\mathbf{Y}\) are the intercept vector and affine-policy
coefficient matrix, respectively. This policy provides
a tractable approximation of unrestricted recourse.

Substituting this policy into the linear cost expression
in \eqref{eq:operation_loss} yields
\begin{equation}
\label{eq:affine_loss_compact}
\begin{aligned}
L_{\mathrm{affine}}(A,\xi)
&= \Phi_0(y)
 + \sum_{k=1}^{K}\Phi_k(\mathbf{Y})\xi_k \\
&= \Phi_0(y)+\Phi(\mathbf{Y})^\top\xi,
\end{aligned}
\end{equation}
where \(A=\{y,\mathbf{Y}\}\) collects the affine recourse
variables and
\(\Phi(\mathbf{Y})
=(\Phi_1(\mathbf{Y}),\ldots,\Phi_K(\mathbf{Y}))^\top\).
Here, \(\Phi_0(y)\) and \(\Phi_k(\mathbf{Y})\) are obtained
by evaluating the linear cost expression in
\eqref{eq:operation_loss} at \(y\) and at the \(k\)-th
column of \(\mathbf{Y}\), respectively.

Because the worst-case distribution is supported on the
empirical atoms, the scenario-wise loss is
\begin{equation}
\label{eq:scenario_loss_affine}
\begin{aligned}
L_s(A)
&= L_{\mathrm{affine}}(A,\widehat{\zeta}_s) \\
&= \Phi_0(y)
 + \Phi(\mathbf{Y})^\top\widehat{\zeta}_s,
\qquad s=1,\ldots,n.
\end{aligned}
\end{equation}
Accordingly, distributional risk is evaluated on the
empirical atoms, while the physical constraints retain
robust feasibility over the full support \(\Xi\).

Under the same affine policy, finite-dimensional
deterministic constraints are obtained as follows.
The operational equalities are enforced by matching
their constant and uncertainty-dependent coefficients:
\begin{subequations}
\label{eq:compact_equalities}
\begin{gather}
A_{\mathrm{eq}}y=b_{\mathrm{eq}},
\\
A_{\mathrm{eq}}\mathbf{Y}=B_{\mathrm{unc}}(x).
\end{gather}
\end{subequations}
Here, \(B_{\mathrm{unc}}(x)\) collects the uncertainty
coefficients and may depend affinely on \(x\) through
\eqref{eq:renewable_output}. The physical inequalities
take the compact robust form
\begin{equation}
\label{eq:compact_robust_inequality}
A_{\mathrm{in}}y
+(A_{\mathrm{in}}\mathbf{Y}-E)\xi
\leq Hx+h,
\qquad \forall \xi\in\Xi.
\end{equation}
Here, \(E\) collects right-hand-side uncertainty
coefficients, including those in the load-shedding
and renewable-curtailment bounds, and may depend
affinely on \(x\) through \eqref{eq:renewable_output}.

Assume the polyhedral uncertainty support
\begin{equation}
\label{eq:polyhedral_support}
\Xi=\{\xi\mid D\xi\leq d\}.
\end{equation}
For row \(m\), let \(a_m^\top\) and \(e_m^\top\) be the
corresponding rows of \(A_{\mathrm{in}}\) and \(E\).
The robust constraint is equivalent to
\begin{equation}
\label{eq:rowwise_robust_constraint}
\max_{\xi\in\Xi}
\left\{(a_m^\top\mathbf{Y}-e_m^\top)\xi\right\}
\leq (Hx)_m+h_m-a_m^\top y.
\end{equation}

\begin{proposition}[Finite robust counterpart]
\label{prop:robust_counterpart}
Suppose that the polyhedral support \(\Xi\) in
\eqref{eq:polyhedral_support} is nonempty and bounded.
For each \(m\in\mathcal{M}_{\mathrm{in}}\), constraint
\eqref{eq:rowwise_robust_constraint} holds if and only if
there exists \(\mu_m\) satisfying
\begin{subequations}
\label{eq:robust_counterpart}
\begin{gather}
d^\top\mu_m
\leq (Hx)_m+h_m-a_m^\top y,
\label{eq:robust_counterpart_a}
\\
D^\top\mu_m
=(a_m^\top\mathbf{Y}-e_m^\top)^\top,
\label{eq:robust_counterpart_b}
\\
\mu_m\geq0,
\qquad m\in\mathcal{M}_{\mathrm{in}}.
\label{eq:robust_counterpart_c}
\end{gather}
\end{subequations}
\end{proposition}

\noindent\textit{Proof.}\quad
See Appendix~\ref{app:robust_counterpart}.

Proposition~\ref{prop:robust_counterpart} provides a finite
deterministic representation of the robust operational
inequalities.

Non-anticipativity permits each decision to depend only on
uncertainties observed by its decision time. Let \(t_m\)
and \(t_k\) denote the time indices of decision row \(m\)
and uncertainty component \(\xi_k\), respectively. Then
\begin{equation}
\label{eq:nonanticipativity}
\mathbf{Y}_{m,k}=0,
\qquad
\forall (m,k)\ \mathrm{s.t.}\ t_k>t_m.
\end{equation}

\subsection{Reformulation of the KLRW-DRO Risk}

For a fixed affine policy \(A=\{y,\mathbf{Y}\}\), the KLRW-DRO
operational risk is
\begin{equation}
\label{eq:klrw_risk_primal}
R_{\mathrm{KLRW}}(A)
=
\sup_{q,T}
\sum_{i=1}^{n}
q_iL_i(A)
\end{equation}
subject to
\begin{equation}
\label{eq:klrw_risk_marginals}
\begin{gathered}
\sum_{j=1}^{n}T_{ij}=q_i,
\qquad
\sum_{i=1}^{n}T_{ij}=p_j,\\
T_{ij}\geq0,
\qquad
q\in\Delta_n,
\end{gathered}
\end{equation}
and
\begin{equation}
\label{eq:klrw_risk_radius}
(1-\beta)
\sum_{i=1}^{n}
\sum_{j=1}^{n}
C_{ij}T_{ij}
+
\beta
\sum_{i=1}^{n}
q_i\log\frac{q_i}{p_i}
\leq
\rho_{\mathrm{KLRW}} .
\end{equation}

\begin{lemma}[Strict feasibility and strong duality]
\label{lem:klrw_strong_duality}
Suppose that \(p_i>0\), \(C_{ij}<\infty\), \(C_{ii}=0\),
\(\beta\in(0,1)\), and \(\rho_{\mathrm{KLRW}}>0\). Then the
finite-dimensional inner problem
\eqref{eq:klrw_risk_primal}--\eqref{eq:klrw_risk_radius}
satisfies Slater's condition relative to its marginal-equality
constraints. Consequently, its primal and dual optimal values
coincide.
\end{lemma}

\noindent\textit{Proof.}\quad
See Appendix~\ref{app:strict_feasibility}.

Lemma~\ref{lem:klrw_strong_duality} yields the following exact
dual reformulation.

\begin{theorem}[Exact dual reformulation of the KLRW-DRO risk]
\label{thm:klrw_risk_dual}
For a fixed affine policy \(A=\{y,\mathbf{Y}\}\) with finite
scenario-wise losses \(L_i(A)\), and under the conditions of
Lemma~\ref{lem:klrw_strong_duality}, the inner risk
\eqref{eq:klrw_risk_primal}--\eqref{eq:klrw_risk_radius}
is equivalent to
\begin{equation}
\label{eq:klrw_risk_dual}
\begin{aligned}
R_{\mathrm{KLRW}}(A)
&=
\inf_{\lambda\geq0,\phi,\psi}\,
\lambda\rho_{\mathrm{KLRW}}
+\sum_{j=1}^{n}p_j\psi_j
\\[-0.3ex]
&\quad
+\lambda\beta
\log\!\left(
\sum_{i=1}^{n}
p_i
\exp\!\left(
\frac{L_i(A)-\phi_i}{\lambda\beta}
\right)
\right).
\end{aligned}
\end{equation}
subject to
\begin{equation}
\label{eq:klrw_transport_dual_constraint}
\phi_i-\psi_j
\leq
\lambda(1-\beta)C_{ij},
\qquad
i,j=1,\ldots,n .
\end{equation}
At \(\lambda=0\), the log-sum-exp perspective in
\eqref{eq:klrw_risk_dual} is interpreted through its
lower-semicontinuous closure, whose value is
\begin{equation}
\label{eq:closed_perspective_limit}
\max_{i=1,\ldots,n}\{L_i(A)-\phi_i\}.
\end{equation}
\end{theorem}

\noindent\textit{Proof.}\quad
See Appendix~\ref{app:klrw_risk_dual}.

Theorem~\ref{thm:klrw_risk_dual} converts the worst-case
expectation into a finite-dimensional convex minimization.

\subsection{Exponential Cone Reformulation}

The log-sum-exp term in \eqref{eq:klrw_risk_dual} admits
an exponential cone representation, derived in
Appendix~\ref{app:exponential_cone}. Let \(\nu\) be its
epigraph variable and \(u=(u_1,\ldots,u_n)^\top\) the
auxiliary variable vector. Combining this representation
with the affine-policy, robust-counterpart, and
transport-dual constraints yields the following
exponential cone program, where \(K_{\exp}\)
is the closed exponential cone defined in
\eqref{eq:kexp_definition}:
\begin{equation}
\label{eq:ecp_objective}
\min_{x,y,\mathbf{Y},\lambda,\phi,\psi,\nu,u,\{\mu_m\}}
\;
I(x)
+\lambda\rho_{\mathrm{KLRW}}
+\sum_{j=1}^{n}p_j\psi_j
+\nu .
\end{equation}
subject to
\begin{subequations}
\label{eq:ecp_constraints}
\begin{gather}
x\in\mathcal{X},
\label{eq:ecp_constraints_a}
\\
A_{\mathrm{eq}}y=b_{\mathrm{eq}},
\qquad
A_{\mathrm{eq}}\mathbf{Y}=B_{\mathrm{unc}}(x),
\label{eq:ecp_constraints_b}
\\
d^\top\mu_m
\leq
(Hx)_m+h_m-a_m^\top y,
\qquad
m\in\mathcal{M}_{\mathrm{in}},
\label{eq:ecp_constraints_c}
\\
D^\top\mu_m
=
(a_m^\top\mathbf{Y}-e_m^\top)^\top,
\quad
\mu_m\geq0,
\quad
m\in\mathcal{M}_{\mathrm{in}},
\label{eq:ecp_constraints_d}
\\
\phi_i-\psi_j
\leq
\lambda(1-\beta)C_{ij},
\qquad
i,j=1,\ldots,n,
\label{eq:ecp_constraints_e}
\\
\sum_{s=1}^{n}u_s
\leq
\lambda\beta,
\label{eq:ecp_constraints_f}
\\
\big(
u_s,\lambda\beta,
L_s(A)-\phi_s-\nu+\lambda\beta\log p_s
\big)
\in
K_{\exp},
\label{eq:ecp_constraints_g}
\\
\mathbf{Y}_{m,k}=0,
\qquad
\forall (m,k)\ \mathrm{s.t.}\ t_k>t_m,
\label{eq:ecp_constraints_h}
\\
\lambda\geq0,
\qquad
u_s\geq0,
\quad
s=1,\ldots,n .
\label{eq:ecp_constraints_i}
\end{gather}
\end{subequations}

\begin{corollary}[Exponential cone representability]
\label{cor:ecp_reformulation}
Under the affine-policy restriction and the assumptions of
Theorem~\ref{thm:klrw_risk_dual} and
Proposition~\ref{prop:robust_counterpart}, the KLRW-DRO
capacity-planning formulation admits the finite-dimensional
convex exponential cone representation
\eqref{eq:ecp_objective}--\eqref{eq:ecp_constraints}.
\end{corollary}

\noindent\textit{Proof.}\quad
See Appendix~\ref{app:exponential_cone}.

\section{Case Study}
In this section, the modified PJM 5-bus and IEEE 118-bus
systems are used to evaluate the planning performance
and computational viability of the proposed KLRW-DRO
approach, respectively. The former examines the effects
of LPEE incorporation and worst-case probability
redistribution, while the latter assesses computational
viability on a larger transmission network.

\subsection{System Configurations}

The case studies use ten years of wind, solar, and load data
from the Open Power System Data platform~\cite{wiese2019open}.
The temporally aligned hourly data are segmented into daily
scenarios and divided chronologically into non-overlapping
training and testing subsets. The training subset is used to
construct the normal-scenario and LPEE-aware empirical
distributions, calibrate the ambiguity radii through bootstrap,
and obtain the capacity plans, while the testing subset is
reserved exclusively for out-of-sample evaluation of the
fixed capacity plans.

The scheduling horizon is \(T=24\) with
\(\Delta t=1~\mathrm{h}\), and the uncertainty vector includes
wind availability, solar availability, and load demand.
Storage charging and discharging efficiencies are both
\(95\%\), with state-of-charge limits of \(10\%\) and \(90\%\).
Wind and PV investment costs range from
\(3.0\times10^6\) to \(4.0\times10^6~\mathrm{CNY/MW}\),
while storage power and energy costs are
\(1.0\times10^6~\mathrm{CNY/MW}\) and
\(1.2\times10^6~\mathrm{CNY/MWh}\), respectively.
Investment costs are annualized using a \(5\%\) discount
rate and a 20-year lifetime, then converted to daily values.

The models are implemented in MATLAB R2021b using
YALMIP~\cite{lofberg2004yalmip} and solved by
MOSEK~\cite{mosektoolbox}. All simulations are performed
on a laptop equipped with an AMD Ryzen 7 5800H CPU
and 16 GB RAM.

\subsection{Benchmark Cases}
Three benchmark cases are defined to isolate the effects
of LPEE incorporation and KL divergence regularization.

Case 1: Base-WDRO, i.e., conventional WDRO based on the
normal-scenario empirical distribution.

Case 2: LPEE-WDRO, i.e., conventional WDRO based on the
LPEE-aware empirical distribution.

Case 3: LPEE-KLRW, i.e., the proposed KLRW-DRO model
based on the LPEE-aware empirical distribution.

Comparing Case 1 and Case 2 identifies the effect of
LPEE incorporation, while comparing Case 2 and Case 3
isolates the effect of KL divergence regularization
under the same LPEE-aware empirical distribution.

\FloatBarrier
\subsection{Modified PJM 5-Bus System}
\begin{figure}[!t]
\centering
\includegraphics[width=0.95\linewidth]{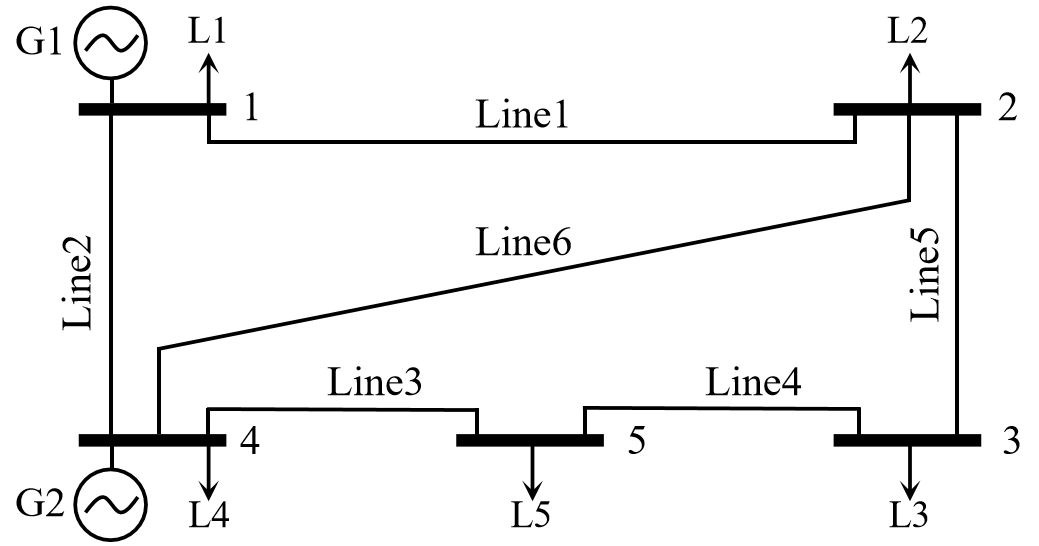}
\caption{Topology of the modified PJM 5-bus system.}
\label{fig:pjm5_topology}
\end{figure}

The modified PJM 5-bus system contains five buses and
six transmission lines, as shown in
Fig.~\ref{fig:pjm5_topology}. Thermal generators are
located at buses 1 and 4, while wind, PV, and storage
are candidate investments at all buses.

\begin{table}[!t]
\caption{Planning Results of the Three Benchmark Cases}
\label{tab:planning_results}
\centering
\footnotesize
\setlength{\tabcolsep}{4.0pt}
\renewcommand{\arraystretch}{1.16}
\setlength{\aboverulesep}{0.45ex}
\setlength{\belowrulesep}{0.45ex}

\begin{tabular*}{\columnwidth}
{@{\hspace{2pt}}c@{\extracolsep{\fill}}ccc@{\hspace{2pt}}}
\toprule
Metric & Case 1 & Case 2 & Case 3 \\
\midrule

Objective (\(10^{6}\) CNY/day)
& 0.7605 & 0.8320 & 0.7173 \\

Dailyized investment (\(10^{6}\) CNY/day)
& 0.1063 & 0.1021 & 0.0879 \\

DRO operational risk (\(10^{6}\) CNY/day)
& 0.6542 & 0.7299 & 0.6294 \\

\addlinespace[0.25em]

Total wind capacity (MW)
& 83.95 & 74.71 & 83.62 \\

Total PV capacity (MW)
& 27.03 & 33.72 & 17.41 \\

Total storage power (MW)
& 67.66 & 68.14 & 63.07 \\

Total storage energy (MWh)
& 64.49 & 55.47 & 27.72 \\

\bottomrule
\end{tabular*}
\vspace{0.35em}
\end{table}

Table~\ref{tab:planning_results} first shows the effect of
incorporating LPEEs into the empirical distribution under
conventional WDRO. From Case 1 to Case 2, the DRO operational
risk increases from \(0.6542\times10^{6}\) to
\(0.7299\times10^{6}~\mathrm{CNY/day}\), whereas the
dailyized investment changes only slightly. Because the two
cases use different empirical distributions and separately
calibrated ambiguity radii, their objective values reflect
different risk representations rather than a direct ranking of
planning performance. Thus, explicitly incorporating LPEEs into
the empirical distribution primarily changes the scarcity risk
represented in the planning model.

Under the same LPEE-aware empirical distribution, the
comparison between Case 2 and Case 3 isolates the effect of KL
divergence regularization. Case 3 reduces the objective by
\(13.78\%\) relative to Case 2, while the DRO operational risk
decreases from \(0.7299\times10^{6}\) to
\(0.6294\times10^{6}~\mathrm{CNY/day}\). The resulting
capacity plan contains more wind capacity and lower PV and
storage capacities, with storage energy decreasing from
\(55.47\) to \(27.72~\mathrm{MWh}\). Therefore, KL divergence
regularization restructures the renewable--storage portfolio
rather than uniformly reducing all capacity investments.

\begin{figure}[!t]
\centering
\includegraphics[width=0.95\linewidth]{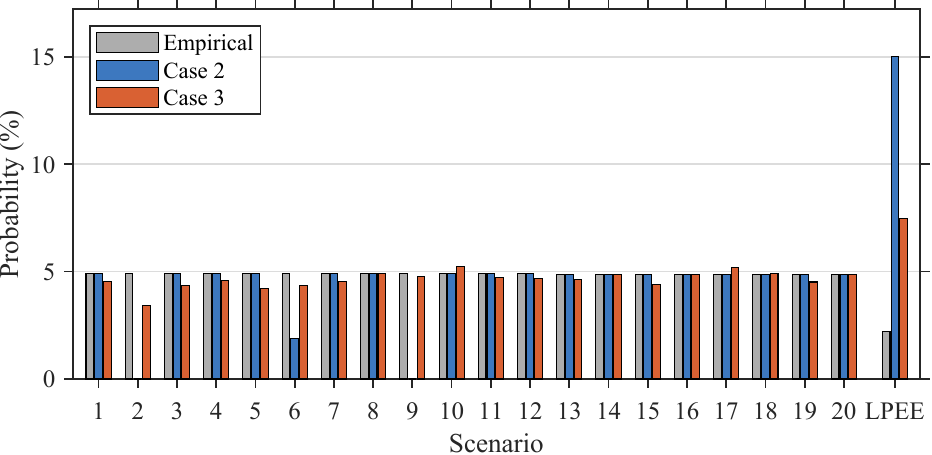}
\caption{Empirical and worst-case probabilities under the
LPEE-aware empirical distribution. Scenario LPEE denotes the
representative LPEE scenario.}
\label{fig:lpee_redistribution}
\end{figure}

The probability redistribution in
Fig.~\ref{fig:lpee_redistribution} further explains the
difference between Case 2 and Case 3. The representative LPEE
scenario has an empirical probability of \(2.2174\%\), which
increases to \(15.0224\%\) under Case 2 and \(7.4634\%\) under
Case 3. The corresponding amplification ratio decreases from
\(6.7748\) to \(3.3658\), while remaining above unity under
Case 3. Overall, the proposed KLRW-DRO model limits excessive
probability amplification without eliminating the risk
information carried by the LPEE scenario.

\subsection{Out-of-Sample Tests}
\label{subsec:oos_tests}

Out-of-sample tests independently evaluate the capacity plans
obtained by Case 2 and Case 3. The investment decisions in
Table~\ref{tab:planning_results} are fixed, and only the
operational variables are re-optimized through deterministic
24-h dispatch for each reserved testing scenario. No
ambiguity-set parameters or capacity decisions are recalibrated,
ensuring that the testing data do not affect planning or
parameter selection.

The testing set separately assesses economic performance under
normal conditions and reliability under LPEE conditions. For
the normal scenarios, the same preprocessing procedure is
applied to both plans to establish a common feasible operating
range, and the economic metric comprises the dailyized
investment cost and the probability-weighted operating cost
conditional on normal scenarios. For the LPEE scenarios, the
original data are retained, and reliability is measured by the
mean expected energy not supplied (EENS) computed using the
prescribed test-set weights. These settings provide a
consistent basis for comparing the two capacity plans.

\begin{table}[!t]
\caption{Out-of-Sample Economic and Reliability Performance}
\label{tab:oos_performance}
\centering
\footnotesize
\setlength{\tabcolsep}{4.0pt}
\renewcommand{\arraystretch}{1.16}
\setlength{\aboverulesep}{0.45ex}
\setlength{\belowrulesep}{0.45ex}

\begin{tabular*}{\columnwidth}
{@{\hspace{2pt}}c@{\extracolsep{\fill}}cc@{\hspace{2pt}}}
\toprule
Metric & Case 2 & Case 3 \\
\midrule

\makecell[c]{Normal OOS economic cost(\(10^{6}\) CNY/day)}
& 0.1598 & 0.1462 \\

\makecell[c]{Mean LPEE OOS EENS(MWh/day)}
& 166.43 & 165.24 \\

\bottomrule
\end{tabular*}
\vspace{0.35em}
\end{table}

Table~\ref{tab:oos_performance} shows that KL divergence
regularization improves out-of-sample economic performance
without increasing EENS under LPEE conditions. Relative to
Case 2, Case 3 reduces the normal OOS economic cost by
approximately \(8.5\%\), from
\(0.1598\times10^{6}\) to
\(0.1462\times10^{6}~\mathrm{CNY/day}\), while the mean
LPEE OOS EENS decreases slightly from \(166.43\) to
\(165.24~\mathrm{MWh/day}\). Thus, Case 3 achieves a more
favorable economy--reliability balance under the reserved
testing scenarios.

\subsection{Sensitivity Analysis}
\label{subsec:sensitivity_analysis}

The sensitivity of KLRW-DRO to the trade-off coefficient
\(\beta\) is examined through the worst-case probability of
the representative LPEE scenario. Because the Wasserstein
distance and KL divergence have different numerical scales, the
effect of \(\beta\) becomes more pronounced as it approaches
one. To better illustrate this region, we define
\(\theta=-\log_{10}(1-\beta)\) and present the sensitivity
results with respect to \(\theta\). For each candidate
\(\beta\), the corresponding
\(\rho_{\mathrm{KLRW}}(\beta)\) is recalibrated using the same
bootstrap replications and the same \(95\%\) quantile level,
while all other model settings remain unchanged.

\begin{figure}[!t]
\centering
\includegraphics[width=\linewidth]{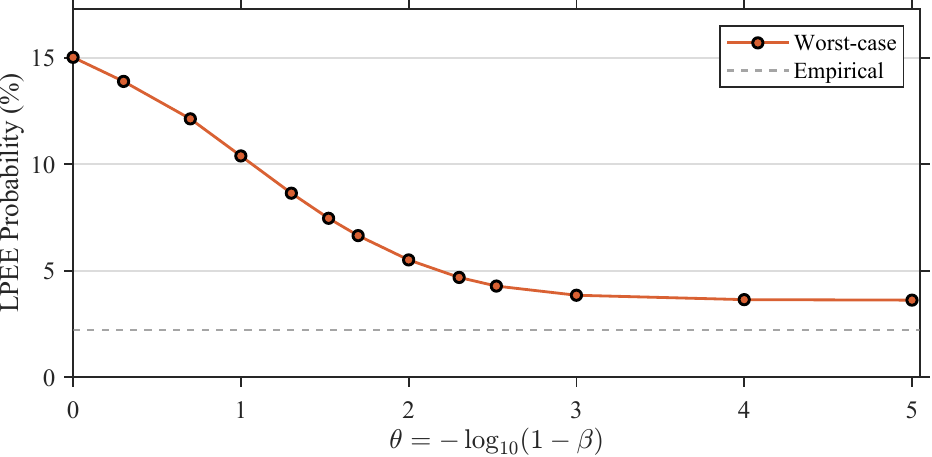}
\caption{Sensitivity of the worst-case LPEE probability to the
trade-off coefficient \(\beta\). The horizontal dashed line
denotes the empirical probability of the representative LPEE
scenario.}
\label{fig:beta_sensitivity}
\end{figure}

As shown in Fig.~\ref{fig:beta_sensitivity}, increasing
\(\beta\) suppresses the worst-case probability of the
representative LPEE scenario. The probability decreases
from \(15.02\%\) at \(\beta=0\) (Case 2) to \(7.46\%\)
at the adopted value \(\beta=0.97\) (Case 3), a reduction
of approximately \(50\%\). Further increases in \(\beta\)
yield diminishing marginal reductions, with the
probability approaching \(3.62\%\) as \(\beta\) approaches
one. Thus, the adopted value substantially limits
excessive probability concentration toward LPEEs without
requiring \(\beta\) to approach its limiting value.

\subsection{Scalability Test on the Modified IEEE 118-Bus System}
\label{subsec:ieee118_scalability}

The modified IEEE 118-bus system is used to assess the
computational viability of the proposed reformulation on a
larger transmission network. To limit the size of the affine
recourse model, shared system-level uncertainty factors, a
simplified affine policy~\cite{lorca2017multistage}, and a
short-memory storage recourse structure
\cite{warrington2016rolling} are adopted. Since the planning
mechanism has been examined using the modified PJM 5-bus
system, this test focuses on solution time and LPEE probability
control.

\begin{table}[!t]
\caption{Computational Results on the Modified IEEE 118-Bus System}
\label{tab:ieee118_scalability}
\centering
\footnotesize
\setlength{\tabcolsep}{4.0pt}
\renewcommand{\arraystretch}{1.16}
\setlength{\aboverulesep}{0.45ex}
\setlength{\belowrulesep}{0.45ex}

\begin{tabular*}{\columnwidth}
{@{\hspace{2pt}}c@{\extracolsep{\fill}}cc@{\hspace{2pt}}}
\toprule
Metric & Case 2 & Case 3 \\
\midrule

\makecell[c]{Worst-case LPEE probability}
& 0.1113 & 0.0224 \\

\makecell[c]{LPEE-scenario amplification ratio}
& 22.2565 & 4.4879 \\

Solution time (s)
& 1309.28 & 3625.55 \\

\bottomrule
\end{tabular*}
\vspace{0.35em}
\end{table}

Table~\ref{tab:ieee118_scalability} shows that both models are
solved to optimality on the larger system. Case 3 requires
\(3625.55\)~s, compared with \(1309.28\)~s for Case 2,
reflecting the additional exponential cone constraints
introduced by KL divergence regularization. Meanwhile, the
LPEE-scenario amplification ratio decreases from \(22.2565\)
to \(4.4879\). Thus, the proposed reformulation remains solvable
in about one hour for offline capacity planning and substantially
reduces LPEE probability amplification on the modified IEEE
118-bus system.

\FloatBarrier

\section{Conclusion}
In this paper, a distributionally robust power system planning model considering LPEEs is established. It is revealed that the classical Wasserstein ambiguity set may lead to over-conservative investment decisions because the probabilities of LPEEs can be disproportionately amplified under the worst-case distribution. To mitigate this issue, a novel KL-regularized Wasserstein ambiguity set is proposed together with a bootstrap-based ambiguity-set radius calibration method. Besides, a tractable solution approach is further developed by employing affine policies and reformulating the problem into an equivalent exponential-cone form through the strong duality theory.

It is expected that this study could provide a better understanding of how LPEEs affect distributionally robust planning and inspire further theoretical and practical works on uncertainty-aware power system planning.

\bibliographystyle{IEEEtran}
\bibliography{mybib}

\begin{appendices}

\section{Proof of Lemma~\ref{lem:support_reduction}}
\label{app:support_reduction}
\setcounter{equation}{0}
\renewcommand\theequation{A.\arabic{equation}}

\noindent\textit{Proof.}\quad
Assume that the empirical distribution is
\begin{equation}
\widehat{\mathbb P}
=
\sum_{s=1}^{n}p_s\delta_{\widehat{\zeta}_s},
\qquad
p_s>0,\quad
\sum_{s=1}^{n}p_s=1.
\end{equation}
For any feasible distribution \(P\) satisfying
\begin{equation}
(1-\beta)W_c(P,\widehat{\mathbb P})
+
\beta D_{\mathrm{KL}}(P\Vert\widehat{\mathbb P})
\leq \rho_{\mathrm{KLRW}} < \infty ,
\end{equation}
the conditions \(W_c(P,\widehat{\mathbb P})\geq0\) and
\(\beta>0\) imply
\begin{equation}
D_{\mathrm{KL}}(P\Vert\widehat{\mathbb P})<\infty.
\end{equation}
By the definition of KL divergence, finite relative entropy implies
\begin{equation}
P\ll \widehat{\mathbb P}.
\end{equation}
Because \(\widehat{\mathbb P}\) is supported on
\(\{\widehat{\zeta}_1,\ldots,\widehat{\zeta}_n\}\), absolute continuity gives
\begin{equation}
P\left(
\Xi\setminus\{\widehat{\zeta}_1,\ldots,\widehat{\zeta}_n\}
\right)=0.
\end{equation}
Hence, any feasible distribution admits the finite-dimensional representation
\begin{equation}
P
=
\sum_{i=1}^{n}q_i\delta_{\widehat{\zeta}_i},
\qquad
q\in\Delta_n .
\end{equation}
This proves the support reduction used in the main formulation.

\section{Proof of Proposition~\ref{prop:robust_counterpart}}
\label{app:robust_counterpart}
\setcounter{equation}{0}
\renewcommand\theequation{B.\arabic{equation}}

\noindent\textit{Proof.}\quad
Consider the \(m\)-th robust inequality
\begin{equation}
a_m^\top y
+
(a_m^\top\mathbf{Y}-e_m^\top)\xi
\leq
(Hx)_m+h_m,
\qquad
\forall \xi\in\Xi,
\end{equation}
where
\begin{equation}
\Xi=\{\xi\mid D\xi\leq d\}.
\end{equation}
It is equivalent to
\begin{equation}
\max_{\xi}
\left\{
(a_m^\top\mathbf{Y}-e_m^\top)\xi:
D\xi\leq d
\right\}
\leq
(Hx)_m+h_m-a_m^\top y.
\end{equation}
The inner maximization is a linear program. Its dual is
\begin{equation}
\min_{\mu_m\geq0}
\left\{
d^\top\mu_m:
D^\top\mu_m
=
(a_m^\top\mathbf{Y}-e_m^\top)^\top
\right\}.
\end{equation}
Since \(\Xi\) is nonempty and bounded, the primal linear program
is feasible and finite. Hence, LP strong duality applies. Therefore,
the robust constraint holds if and only if there exists
\(\mu_m\geq0\) such that
\begin{equation}
d^\top\mu_m
\leq
(Hx)_m+h_m-a_m^\top y,
\end{equation}
and
\begin{equation}
D^\top\mu_m
=
(a_m^\top\mathbf{Y}-e_m^\top)^\top.
\end{equation}
Applying this argument to all \(m\in\mathcal{M}_{\mathrm{in}}\)
yields the finite deterministic robust counterpart.

\section{Proof of Lemma~\ref{lem:klrw_strong_duality}}
\label{app:strict_feasibility}
\setcounter{equation}{0}
\renewcommand\theequation{C.\arabic{equation}}

\noindent\textit{Proof.}\quad
Consider the candidate point
\begin{equation}
\begin{aligned}
T^\varepsilon
&=
(1-\varepsilon)\operatorname{diag}(p)
+
\varepsilon pp^\top,\\
q^\varepsilon
&=
p,\qquad
\varepsilon\in(0,1).
\end{aligned}
\end{equation}
Since \(p_i>0\) and \(\varepsilon\in(0,1)\), we have
\(q_i^\varepsilon>0\) and \(T^\varepsilon_{ij}>0\) for all
\(i,j\). Moreover, for each \(i\),
\begin{equation}
\sum_{j=1}^{n}T^\varepsilon_{ij}
=
(1-\varepsilon)p_i
+
\varepsilon p_i\sum_{j=1}^{n}p_j
=
p_i
=
q_i^\varepsilon,
\end{equation}
and for each \(j\),
\begin{equation}
\sum_{i=1}^{n}T^\varepsilon_{ij}
=
(1-\varepsilon)p_j
+
\varepsilon p_j\sum_{i=1}^{n}p_i
=
p_j.
\end{equation}
Thus, \(T^\varepsilon\) has the required marginals \((p,p)\).
Since \(q^\varepsilon=p\),
\begin{equation}
D_{\mathrm{KL}}(q^\varepsilon\Vert p)
=
D_{\mathrm{KL}}(p\Vert p)
=
0.
\end{equation}
Using \(C_{ii}=0\), we have
\begin{equation}
\sum_{i,j}C_{ij}T^\varepsilon_{ij}
=
\varepsilon
\sum_{i,j}p_ip_jC_{ij}.
\end{equation}
Let
\begin{equation}
M
:=
\sum_{i,j}p_ip_jC_{ij}.
\end{equation}
Since \(C_{ij}<\infty\), \(M<\infty\). If \(M>0\), choose
\begin{equation}
0<\varepsilon
<
\min\left\{
1,
\frac{\rho_{\mathrm{KLRW}}}{(1-\beta)M}
\right\}.
\end{equation}
If \(M=0\), any \(\varepsilon\in(0,1)\) is admissible. Hence, for
\(\rho_{\mathrm{KLRW}}>0\),
\begin{equation}
(1-\beta)
\sum_{i,j}C_{ij}T^\varepsilon_{ij}
+
\beta D_{\mathrm{KL}}(q^\varepsilon\Vert p)
<
\rho_{\mathrm{KLRW}}.
\end{equation}
Therefore, the KLRW-DRO ambiguity-radius constraint is strictly
satisfied, while \((q^\varepsilon,T^\varepsilon)\) lies in the
relative interior of the nonnegative domain subject to the marginal
equalities. Slater's condition holds relative to the affine hull of
the marginal constraints, and thus there is no duality gap.

\section{Proof of Theorem~\ref{thm:klrw_risk_dual}}
\label{app:klrw_risk_dual}
\setcounter{equation}{0}
\renewcommand\theequation{D.\arabic{equation}}

\noindent\textit{Proof.}\quad
By Lemma~\ref{lem:support_reduction}, the inner KLRW-DRO risk can be written as
\begin{equation}
R_{\mathrm{KLRW}}(A)
=
\sup_{q\in\Delta_n,\;T\geq0}
\sum_{i=1}^{n}q_iL_i(A)
\end{equation}
subject to
\begin{equation}
\sum_{j=1}^{n}T_{ij}=q_i,\quad
\sum_{i=1}^{n}T_{ij}=p_j,
\end{equation}
and
\begin{equation}
(1-\beta)
\sum_{i=1}^{n}\sum_{j=1}^{n}C_{ij}T_{ij}
+
\beta\sum_{i=1}^{n}q_i\log\frac{q_i}{p_i}
\leq \rho_{\mathrm{KLRW}} .
\end{equation}

Let \(\lambda\geq0\) be the dual variable associated with the KLRW-DRO ambiguity-radius constraint, and let \(\widetilde{\phi}_i\in\mathbb R\) and \(\psi_j\in\mathbb R\) be the dual variables associated with \(\sum_jT_{ij}=q_i\) and \(\sum_iT_{ij}=p_j\), respectively. The Lagrangian is
\begin{align}
\mathcal{L}
={}&
\sum_i q_iL_i(A)
-\lambda\Biggl(
(1-\beta)\sum_{i,j}C_{ij}T_{ij}
\notag\\[-0.3ex]
&\qquad
+\beta\sum_i q_i\log\frac{q_i}{p_i}
-\rho_{\mathrm{KLRW}}
\Biggr)
-\sum_i\widetilde{\phi}_i
\notag\\[-0.3ex]
&\qquad
\left(
\sum_jT_{ij}-q_i
\right)
-\sum_j\psi_j
\left(
\sum_iT_{ij}-p_j
\right).
\label{eq:lagrangian}
\end{align}
Collecting the terms associated with \(q\) and \(T\) gives
\begin{align}
\mathcal{L}
={}&
\lambda\rho_{\mathrm{KLRW}}
+\sum_j p_j\psi_j
+\sum_i q_i
\left(L_i(A)+\widetilde{\phi}_i\right)
\notag\\[-0.3ex]
&-
\lambda\beta
\sum_i q_i\log\frac{q_i}{p_i}
+
\sum_{i,j}T_{ij}
\notag\\[-0.3ex]
&\qquad\times
\left(
-\widetilde{\phi}_i-\psi_j
-\lambda(1-\beta)C_{ij}
\right).
\label{eq:lagrangian_reform}
\end{align}
The supremum over \(T\geq0\) is finite if and only if
\begin{equation}
-\widetilde{\phi}_i-\psi_j
-\lambda(1-\beta)C_{ij}\leq0,
\qquad \forall i,j.
\end{equation}
Let
\begin{equation}
\phi_i=-\widetilde{\phi}_i.
\end{equation}
Then the above condition becomes
\begin{equation}
\phi_i-\psi_j\leq \lambda(1-\beta)C_{ij},
\qquad \forall i,j.
\end{equation}

The remaining \(q\)-block is
\begin{equation}
\sup_{q\in\Delta_n}
\left\{
\sum_iq_i(L_i(A)-\phi_i)
-
\lambda\beta
\sum_iq_i\log\frac{q_i}{p_i}
\right\}.
\end{equation}
For \(\lambda\beta>0\), the convex conjugate of relative entropy gives
\begin{align}
&
\sup_{q\in\Delta_n}
\left\{
\sum_iq_i a_i
-
\lambda\beta
\sum_iq_i\log\frac{q_i}{p_i}
\right\}
\notag\\
&\qquad =
\lambda\beta
\log
\left(
\sum_i p_i
\exp\left(\frac{a_i}{\lambda\beta}\right)
\right),
\end{align}
where \(a_i=L_i(A)-\phi_i\). Therefore,
\begin{align}
R_{\mathrm{KLRW}}(A)
={}&
\inf_{\lambda\geq0,\phi,\psi}\,
\lambda\rho_{\mathrm{KLRW}}
+\sum_{j=1}^{n}p_j\psi_j
+\lambda\beta\log
\notag\\[-0.3ex]
&\qquad
\Biggl(
\sum_{i=1}^{n}p_i
\exp\!\left(
\frac{L_i(A)-\phi_i}{\lambda\beta}
\right)
\Biggr).
\end{align}
subject to
\begin{equation}
\phi_i-\psi_j\leq \lambda(1-\beta)C_{ij},
\qquad \forall i,j.
\end{equation}
At \(\lambda=0\), the closed-perspective limit equals
\(\max_i\{L_i(A)-\phi_i\}\). Strong duality follows from
Lemma~\ref{lem:klrw_strong_duality}.


\section{Exponential Cone Reformulation}
\label{app:exponential_cone}
\setcounter{equation}{0}
\renewcommand\theequation{E.\arabic{equation}}

The remaining nonlinearity in \eqref{eq:klrw_risk_dual} is the
log-sum-exp term. For \(\lambda>0\), introduce an epigraph
variable \(\nu\) such that
\begin{equation}
\label{eq:lse_epigraph}
\begin{aligned}
\nu
\geq
\lambda\beta
\log
\left(
\sum_{s=1}^{n}
p_s
\exp
\left(
\frac{L_s(A)-\phi_s}{\lambda\beta}
\right)
\right).
\end{aligned}
\end{equation}
Equivalently,
\begin{equation}
\label{eq:lse_exponential_form}
\sum_{s=1}^{n}
p_s
\exp
\left(
\frac{L_s(A)-\phi_s-\nu}{\lambda\beta}
\right)
\leq
1 .
\end{equation}
Multiplication by \(\lambda\beta\) gives the perspective
form
\begin{equation}
\label{eq:lse_perspective}
\sum_{s=1}^{n}
\lambda\beta
\exp
\left(
\frac{
L_s(A)-\phi_s-\nu+\lambda\beta\log p_s
}
{\lambda\beta}
\right)
\leq
\lambda\beta .
\end{equation}
Using the exponential cone
\begin{equation}
\label{eq:kexp_definition}
\begin{aligned}
K_{\exp}
=
\operatorname{cl}
\big\{
&(x_1,x_2,x_3)\in\mathbb{R}^3:
\\
&x_1\geq x_2\exp(x_3/x_2),\ x_2>0
\big\}.
\end{aligned}
\end{equation}
the perspective constraint is represented by auxiliary
variables \(u_s\geq0\):
\begin{subequations}
\label{eq:exp_cone_constraints}
\begin{gather}
\sum_{s=1}^{n}
u_s
\leq
\lambda\beta,
\label{eq:exp_cone_sum}
\\
\left(
u_s,\lambda\beta,
L_s(A)-\phi_s-\nu+\lambda\beta\log p_s
\right)
\in
K_{\exp}.
\label{eq:exp_cone_each}
\end{gather}
\end{subequations}
The closed-cone formulation allows \(\lambda=0\).
At this boundary, \eqref{eq:exp_cone_sum} and
\(u_s\geq0\) imply \(u_s=0\), and
\eqref{eq:exp_cone_each} reduces to
\(\nu\geq\max_s\{L_s(A)-\phi_s\}\), consistent with
\eqref{eq:closed_perspective_limit}.

Combining \eqref{eq:exp_cone_constraints} with the
transport-dual constraints and the deterministic
operational constraints yields the finite-dimensional
convex exponential cone program
\eqref{eq:ecp_objective}--\eqref{eq:ecp_constraints}.
This proves Corollary~\ref{cor:ecp_reformulation}.

For numerical stability, the implementation imposes
\begin{equation}
\label{eq:lambda_lower_bound}
\lambda\geq\underline{\lambda}>0.
\end{equation}
This numerical restriction may alter the optimal value
and is not part of the theoretical formulation
\eqref{eq:ecp_objective}--\eqref{eq:ecp_constraints}.

\end{appendices}

\end{document}